\documentclass[11pt,a4paper,twoside]{article}
\usepackage[dvips]{graphics}
\usepackage{cite}
\usepackage{epsfig}
\begin{document}
\setlength{\footskip}{30pt}
\setlength{\oddsidemargin}{0mm}
\setlength{\evensidemargin}{0mm}
\setlength{\topmargin}{0mm}
\setlength{\headheight}{0mm}
\setlength{\headsep}{0mm}
\setlength{\textheight}{230mm}
\setlength{\textwidth}{161mm}
\setlength{\marginparwidth}{0mm}
\setlength{\marginparsep}{0mm}
\clearpage
%
%
\begin{titlepage}
\vspace*{-10mm}
\hbox to \textwidth{ \hsize=\textwidth
\hspace*{0pt\hfill} 
\vbox{ \hsize=58mm
{
\hbox{ PITHA 99/40\hss}
\hbox{ MPI-PhE/99-17 \hss}
\hbox{ December 20, 1999\hss } 
}
}
}

\bigskip\bigskip\bigskip
\begin{center}
{\Huge\bf
Determination of $\mathbf{\alpha_S}$ at $\mathbf{500}$~GeV\\[1.5mm]
from Event Shapes and Jet Rates 
}

\end{center}
\bigskip\bigskip
\begin{center}
O.~Biebel$^{(*)}$
\end{center}
\bigskip

%
\begin{abstract}
\noindent 
The potential of the TESLA linear e$^+$e$^-$ collider to determine the 
strong coupling constant, $\alpha_S$, at $500$~GeV is investigated. 
Experimental complications due to background from W- and Z-pairs, 
top-production, initial state photon radiation and from beamstrahlung 
are considered. The hadronic event selection procedures used by the 
experiments at LEP II are reviewed for the applicability at TESLA. An 
estimate of the various error contributions to the total uncertainty 
of an $\alpha_S$ determination is presented. It confirms that hadronisation 
effects are diminished while the uncertainty from the choice of the 
renormalisation scale will dominate. Fits of the $\ln(R)$-matched 
second order (${\cal O}(\alpha_S^2)$) and resummed calculation (NLLA) 
to six observables are used to estimate the error contributions. This 
yields the expectation of the precision for $\alpha_S(500 {\mathrm{~GeV}})$ 
of $\pm 0.0025$.
%
%
\end{abstract}

\vspace*{0pt\vfill}
\vfill
%
\bigskip\bigskip\bigskip\bigskip

{
\small
\noindent
$^{(*)}$ 
\begin{minipage}[t]{155mm} 
III. Physikalisches Institut der RWTH Aachen, D-52056 Aachen, Germany \\
now at: Max-Planck-Institut f\"ur Physik, D-80805 M\"unchen, Germany \\
contact e-mail: Otmar.Biebel@CERN.ch
\end{minipage}
\hspace*{0pt\hfill}
}

\end{titlepage}
%
%
\newpage
\section{ Introduction }
\label{sec-intro}
The discovery of asymptotic freedom is an important corner stone 
of the foundation of Quantum Chromodynamics (QCD) as the theory of 
the strong interaction. Despite the substantial effort in testing
the predictions of this theory at e$^+$e$^-$ colliders like e.g.\
PETRA~\cite{bib-Maettig-habil} and LEP~\cite{bib-Hebbeker-habil
}, QCD is less precisely surveyed 
than the electroweak sector of the standard model of electroweak 
and strong interaction.  The precision of the determination of the 
strong coupling constant, $\alpha_S$, at centre-of-mass energies of 
up to $200$~GeV at LEP II is only at the level of a few percent 
(see e.g.~\cite{bib-Bethke-world-average}). 

  To test the details of the running of $\alpha_S$, which means the 
asymptotic freedom, one has to strive for a precision 
of better than one percent. This could be achieved by a determination 
of $\alpha_S$ at a linear collider operating at a centre-of-mass energy 
of $500$ GeV or above. It would provide a long lever arm for the test 
of the asymptotic freedom.  

  Besides the running of the coupling constant, QCD also predicts
the value of the renormalised quark masses to depend on the scale of
the interaction process~\cite{bib-Bernreuther-etal
}.
First investigations of this running of the quark masses were performed by 
the LEP and SLC experiments~\cite{bib-L3-PLB271-461},
which suffered from large statistical 
uncertainties. Further experimental manifestation of the effect is, therefore, 
still required. This could be achieved at a linear collider due to its high 
luminosity and by using very high resolution silicon vertex detectors.
Here the silicon micro vertex detectors allow to distinguish bottom and charm
quark flavours created by the annihilation process from the other lighter 
quarks. 

  The identification of the primary quark flavour could furthermore be used
to study the properties of quark and gluon jets. Such studies have already
be successfully conducted at LEP I and SLC (see e.g.~\cite{bib-Gary}).
At the higher energies of LEP II, however, the data statistics is scarce
for a significant investigation of the quark and gluon properties. Again, a 
linear collider would allow such studies because of 
its high specific luminosity.

  Apart from the advantage of very high luminosity, QCD studies at a linear
collider also gain from the high centre-of-mass energy. This is because the
impact of hadronisation, which blurs the view at the quarks and gluons, 
diminishes with a reciprocal power of the scale of the process. Hence the
distortion due to hadronisation effects at $500$~GeV is only about $20\%$
of that at the Z mass scale~\cite{bib-Bethke-Linear-Collider}.
  
  This article presents a scrutiny of the potential of the TESLA linear 
collider to perform a determination of the strong coupling constant at
a centre-of-mass energy of $500$~GeV.

\section{Initial considerations}
\label{sec-basis}
The study is restricted to an investigation based on Monte Carlo 
generator simulations using the PYTHIA 5.722 \cite{bib-PYTHIA}, 
HERWIG 5.9 \cite{bib-HERWIG}, and ARIADNE 4.8 \cite{bib-ARIADNE} 
programs which, except for HERWIG, have been tuned by the OPAL 
collaboration at LEP~\cite{bib-OPAL-ZPC69-543}. 
These generators are used to simulate the production of 
quark-antiquark pairs from the annihilation. To account for the 
substantial production cross-section of massive gauge boson pairs 
(see e.g.~\cite{bib-Boudjema-9809220}), background from hadronic 
decays of Z and W bosons is considered, where in addition also 
the decay W$\rightarrow \tau \nu_{\tau}$ is included. 
All other leptonic decays of these massive bosons can easily 
be identified and, hence, rejected.  

  Likewise, background contributions from $\gamma\gamma$ scattering 
and $\tau$ pair production are not included. Their contribution can
be removed very efficiently as is known from the experiments
at LEP II (see e.g.~\cite{bib-OPAL-ZPC75-193}).

\begin{figure}
\vspace*{-5mm}
\centerline{\epsfxsize=75mm \epsfbox{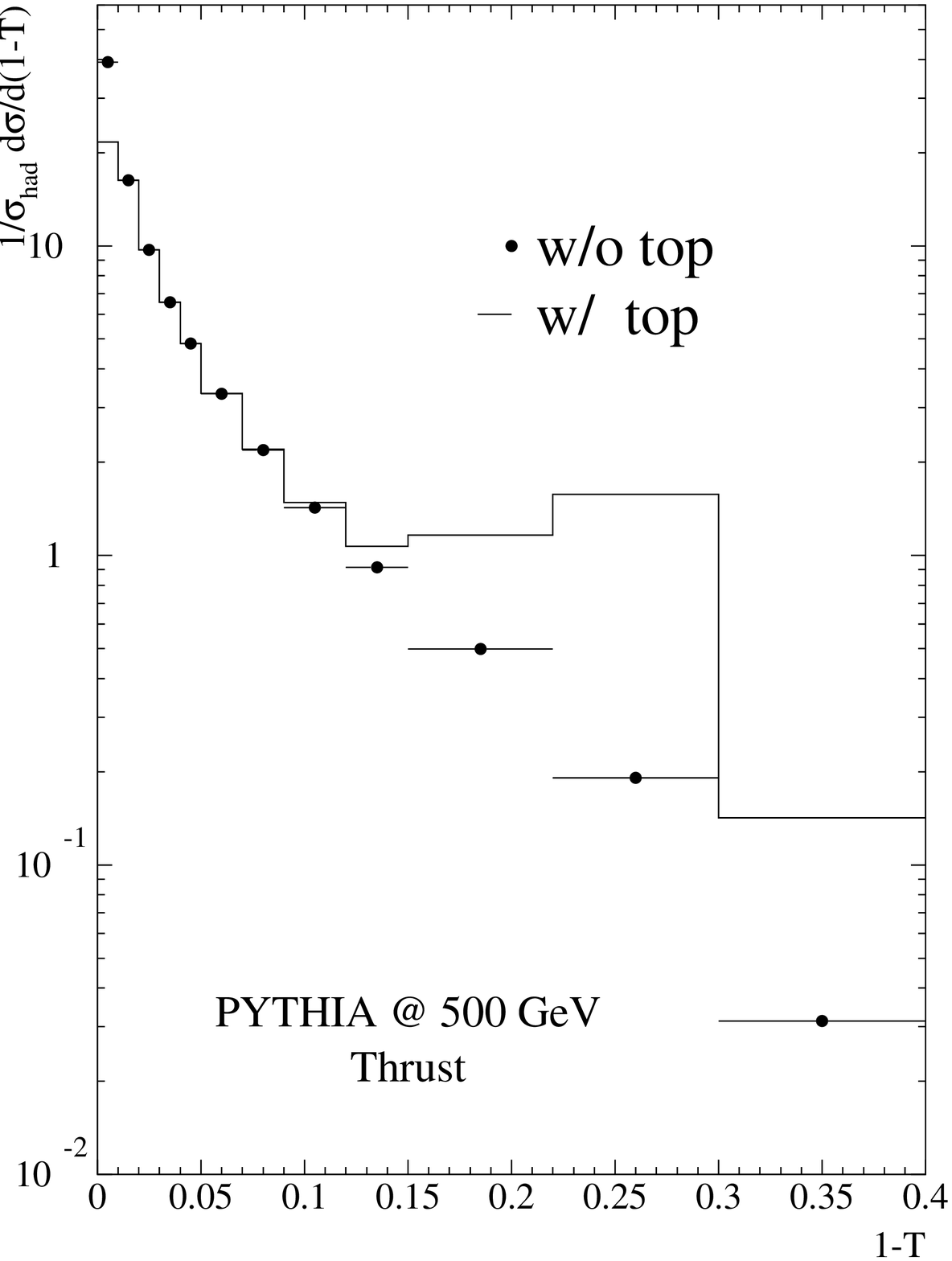}
            \hspace*{5mm}
            \epsfxsize=75mm \epsfbox{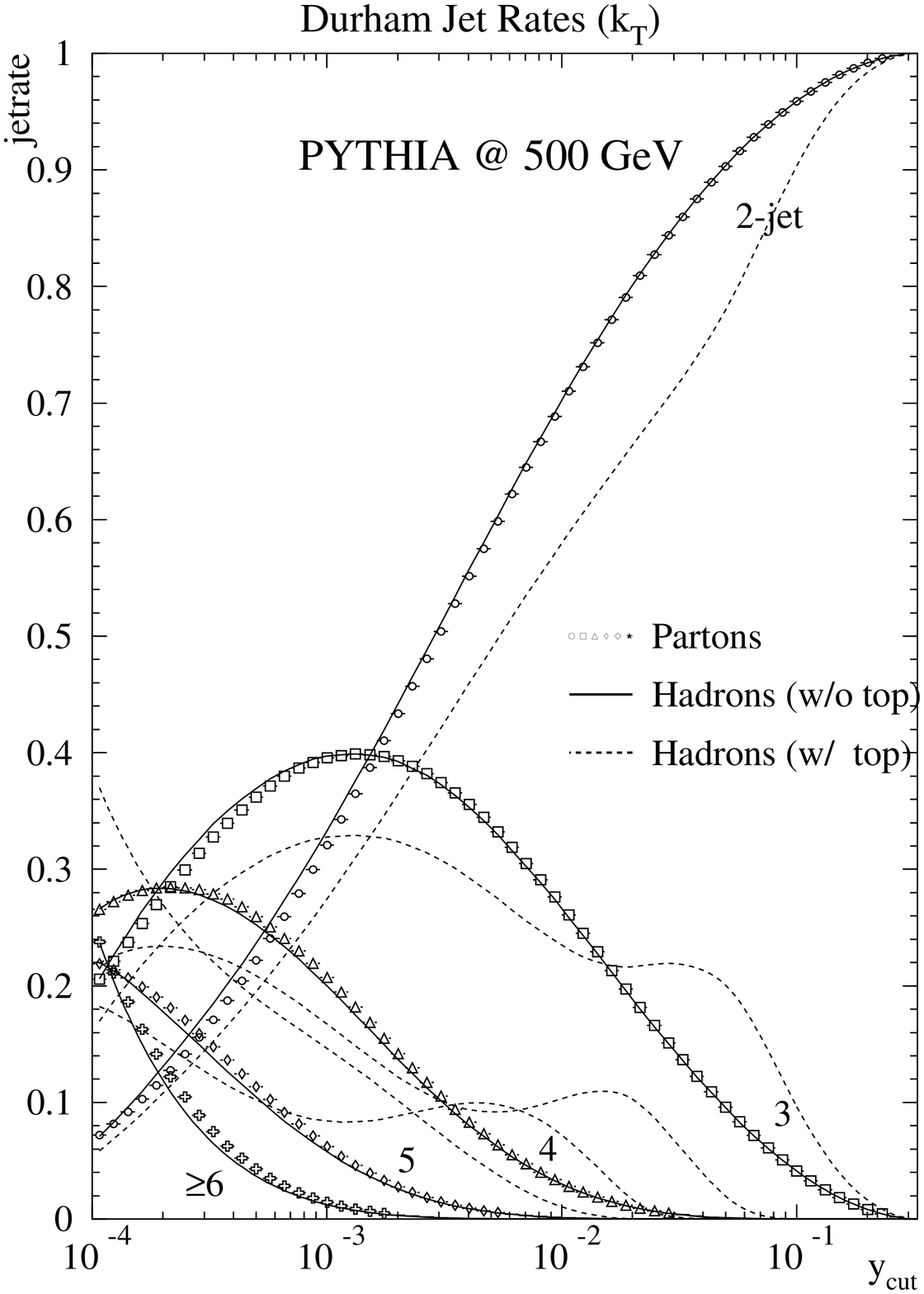}}
\caption{\label{fig-top-distortions}
         Distortions of thrust distribution (left) and of the jet rates 
         (right) due to the top quark pair production. The negligible 
         impact of hadronisation is made visible for the jet rates.
        }
\end{figure}

  The production of top quarks leads to significant effects. This is
due to its large mass and the appearance of typically three jets of 
particles from its decay. Top quark events, therefore, appear rather
spherically in the detector, thus immitating multijet events.
Figure~\ref{fig-top-distortions} shows the impact of the pair-produced 
top quarks on the thrust observable and the jet rates, both 
of which are usually used to determine the value of the strong coupling 
constant.  
It is therefore indispensable to exclude the contribution from top quarks
when performing an $\alpha_S$ determination.

\begin{figure}
\vspace*{-5mm}
\centerline{\epsfxsize=75mm \epsfbox{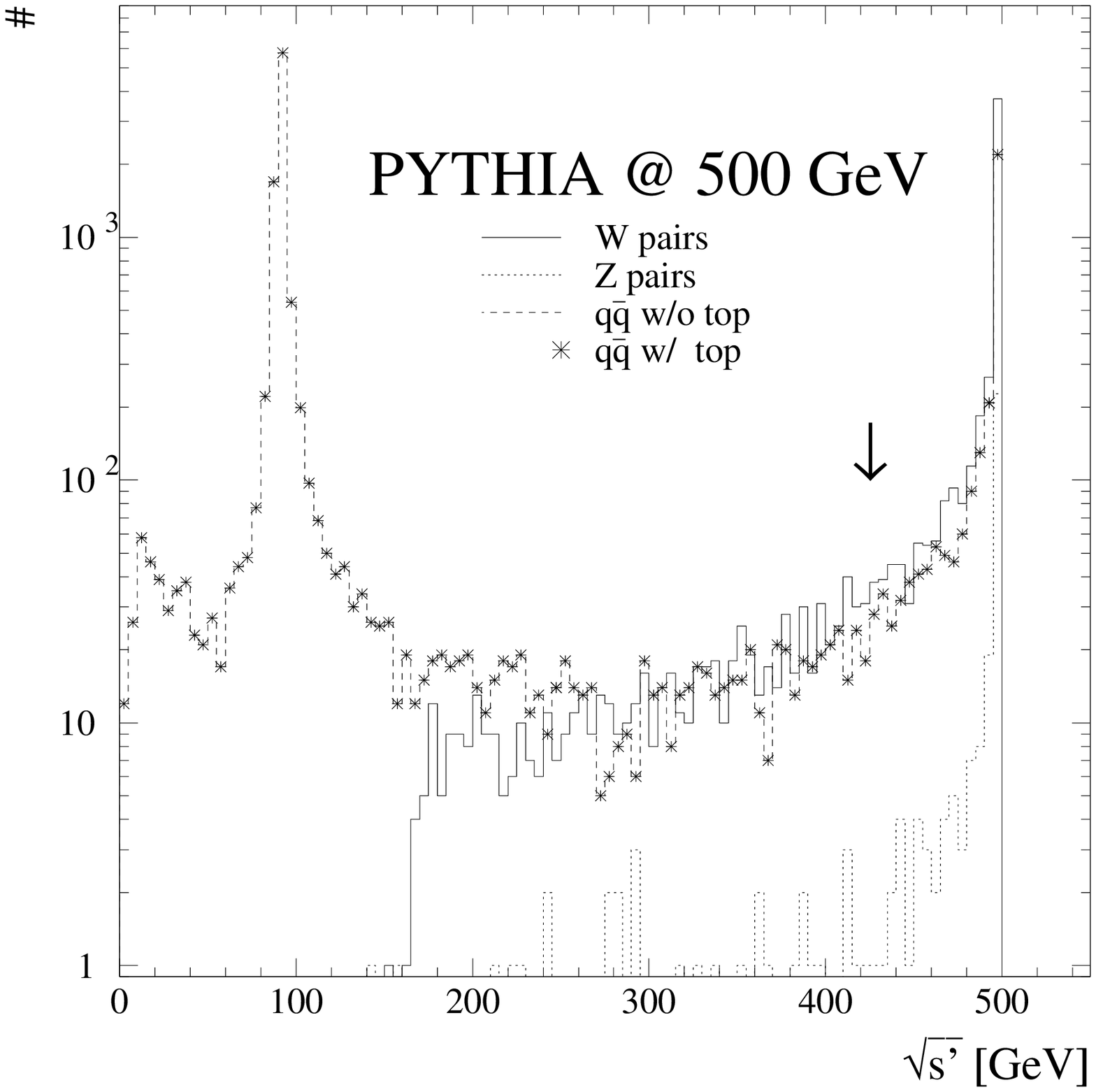}
            \hspace*{5mm}
            \epsfxsize=75mm \epsfbox{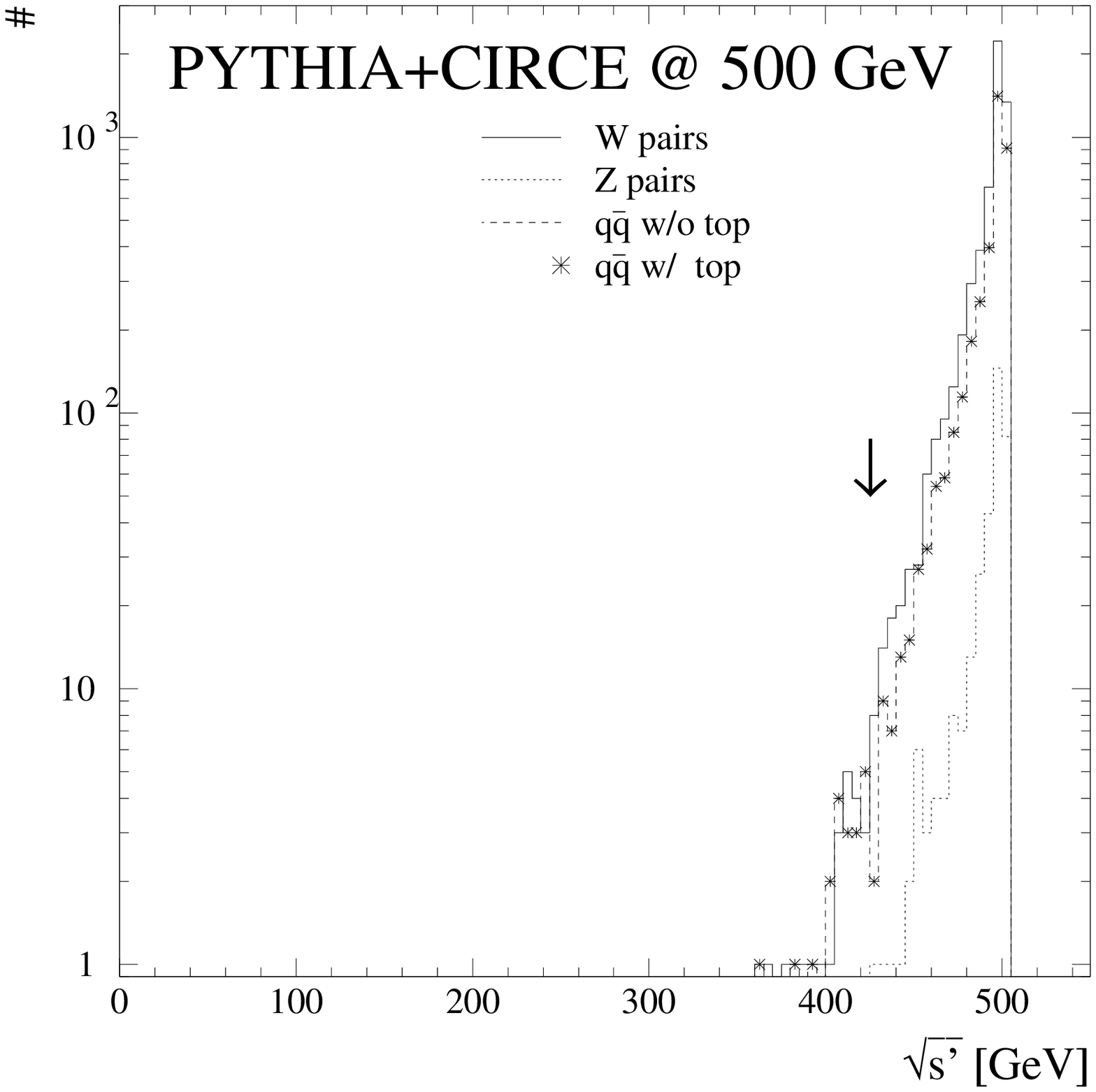}}
\caption{\label{fig-isr+bs}
         Distribution of the effective centre-of-mass energy $\sqrt{s'}$
         affected by initial state radiation (ISR, left) and by beamstrahlung
         (BS, right).
        }
\end{figure}
  The impact of initial state radiation is accounted for using its
next-to-leading order treatment in the PYTHIA generator. Radiative events
constitute a large fraction of the hadronic final states, especially
radiative returns to the Z pole. Most of these can easily be 
rejected by requiring the reconstructed mass of the whole event to be
close to the centre-of-mass energy of the colliding beams. Such a cut
on the effective centre-of-mass energy, $\sqrt{s'}$, is usually applied 
in the hadronic event selection at LEP II (e.g.~\cite{bib-OPAL-ZPC72-191}). 
It effectively takes out radiative Z return events as can be seen in
Figure~\ref{fig-isr+bs}.

  Beamstrahlung, which is of relevance at a linear collider only, 
is simulated using the program CIRCE 1.28\footnote{Parameter set version 5,
revision 1998 05 05}~\cite{bib-CIRCE} which 
has been interfaced to the PYTHIA generator. When using the parameters 
proposed for the TESLA linear collider, only a negligible fraction of the 
events is lost due to beamstrahlung (see Figure~\ref{fig-isr+bs}) if the 
cut $\sqrt{s'}>0.85\cdot\sqrt{s}$ is applied which is usually used by the 
LEP experiments.

\section{Hadronic event selection}
\label{sec-selection}
To study the prospects of a determination of $\alpha_S$ at the TESLA
linear collider, an integrated luminosity of $300$ fb$^{-1}$ has been 
simulated.  This corresponds to roughly one year of data taking at the 
design luminosity of the collider. 
\begin{table}
\begin{center}
\begin{tabular}{|l||c|c|c|c|c|}
\cline{2-6}
       \multicolumn{1}{l|}{}
     & $\mathrm{e^+e^-\rightarrow q\bar{q}}$
     & $\mathrm{e^+e^-(\gamma)\rightarrow q\bar{q}}$
     & $\mathrm{\rightarrow W^+W^-}$
     & $\mathrm{\rightarrow Z Z}$
     & $\mathrm{\rightarrow t\bar{t}} $
\\ \cline{2-6}\hline 
$\sigma(500{\mathrm{~GeV}})$ [pb]
     & $2.8$          & $11.6$         & $4.8$          & $0.3$         & $0.3$
\\ \hline 
\# events
     & $\approx 900$k & $\approx 3.6$M & $\approx 1.5$M & $\approx 90$k & $\approx 90$k
\\ \hline 
\end{tabular}
\end{center}
\caption{\label{tab-no-of-events} 
         Approximate number of events for $300$~fb$^{-1}$ and the cross-sections
         taken from PYTHIA~\protect\cite{bib-PYTHIA}.
        }
\end{table}
Adopting the cross-sections from the PYTHIA program, this integrated 
luminosity translates into the number of events listed in 
Table~\ref{tab-no-of-events}.

  The main issue of the hadronic event selection is to reject the
contributions from W and Z pair, and from top-antitop production.
The option of polarising the electron beam at the TESLA collider,
which allows to strongly suppress the production of W pairs, is 
disregarded for this study. 

  Quite some experience in rejecting W pair events has been collected 
by the experiments at LEP II~(see e.g.~\cite{bib-OPAL-ZPC75-193,
bib-DELPHI-ZPC73-229}).
The detailed examination showed, however, that 
cuts on $4$-jet matrix element weight, $W_4$, and on the $3$- to $4$-jet 
flip $y_{\mathrm{cut}}$ value, $y_{34}$, 
both of which have a good separation
power at LEP II energies, cannot discriminate between W pair and
$q\bar{q}$ events at $500$~GeV. 

\begin{figure}
\vspace*{-5mm}
\centerline{\epsfxsize=75mm \epsfbox{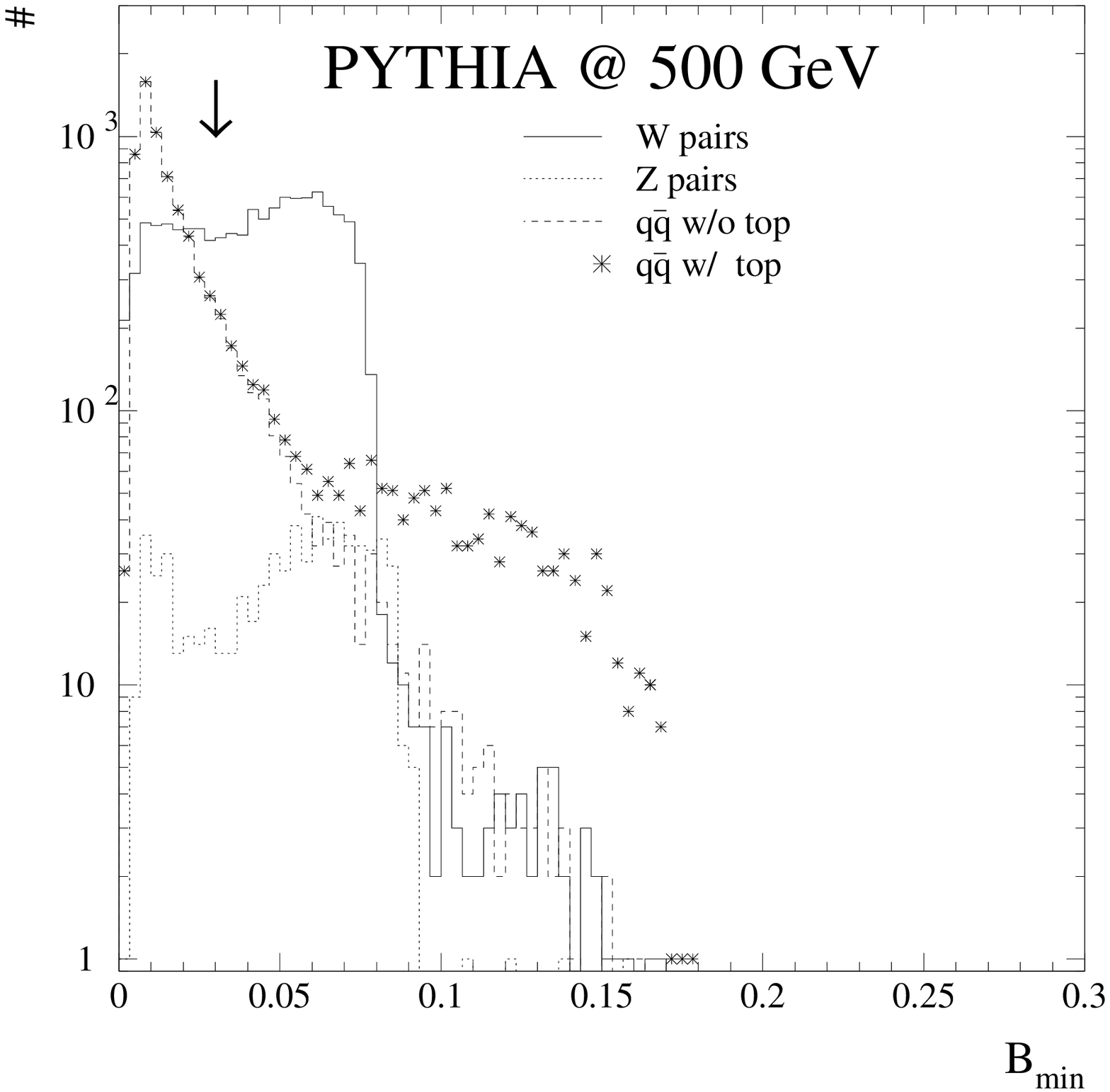}
            \hspace*{5mm}
            \epsfxsize=75mm \epsfbox{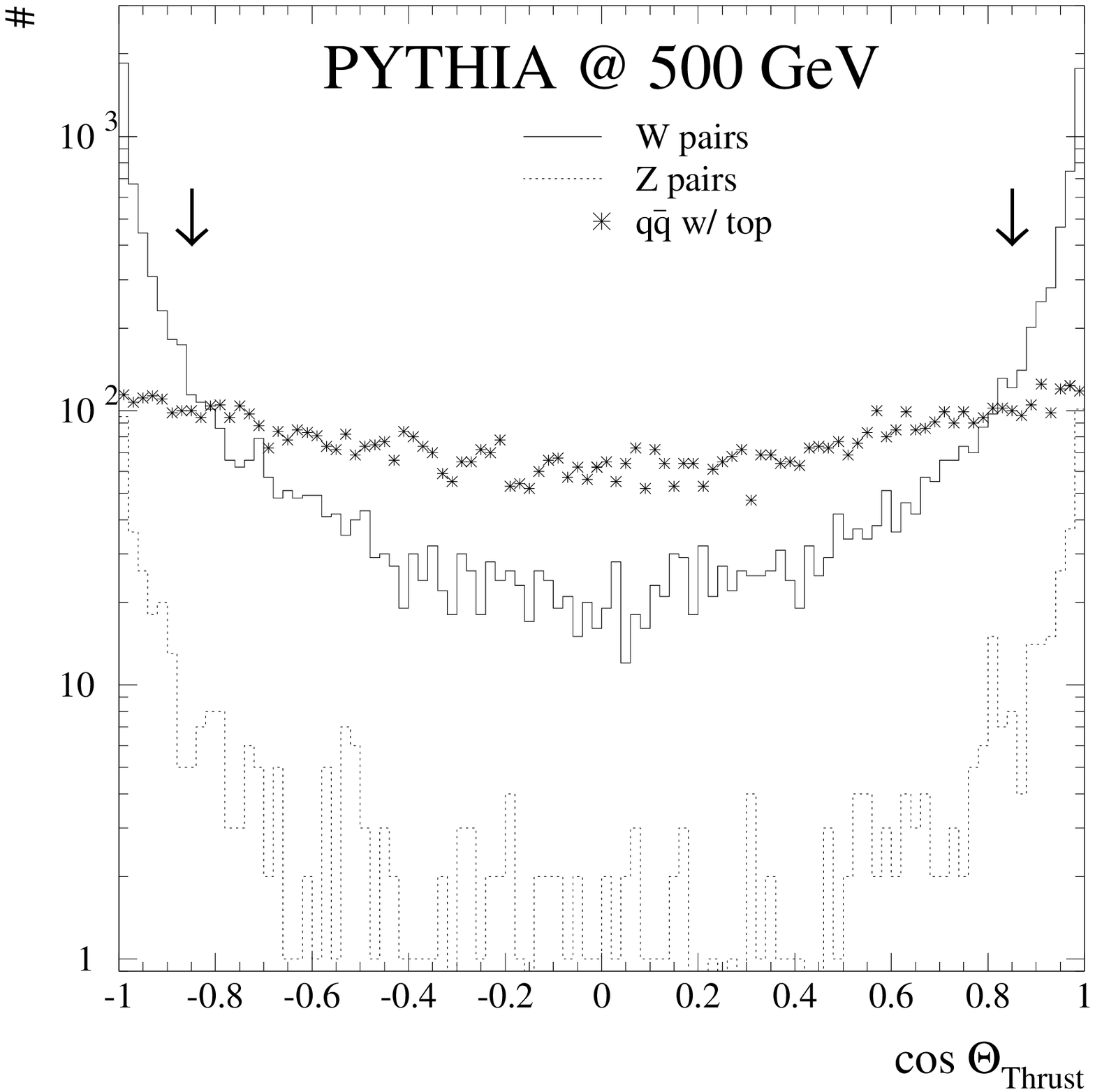}}
\caption{\label{fig-bmin+cos_thrust}
         Effective suppression of W and Z pair, and of top-antitop 
         events by cutting on the minimum value of the jet broadening 
         ($B_{\mathrm{min}}<0.03$, left) and on polar angle of the 
         thrust axis ($|\cos\theta_{\mathrm{thrust}}|<0.85$, right).
        }
\end{figure}

  The minimum value of the jet broadening, $B_{\mathrm{min}} \equiv 
\min(B_1, B_2)$, determined from the jet broadening in each hemisphere 
defined by the thrust axis of an event, proves to powerfully reject
top-antitop events (see Figure~\ref{fig-bmin+cos_thrust}). 
It also reduces the contribution of W and Z pair events. 

  These boson pair events can be eliminated more effectively by cuts 
on the mass of each thrust hemisphere. This is due to the high 
centre-of-mass energy giving both of the bosons a significant boost 
such that the thrust axis yields the vector bosons' direction of flight. 
Thus the hemisphere masses are close to the mass of the W or Z 
boson which allows to easily reject such events. Requiring a minimum 
hemisphere mass of $1\%$ of the centre-of-mass energy eliminates events 
with a W$\rightarrow \tau\nu_{\tau}$ decay. Fully hadronic decays of the 
vector bosons are sorted out if either the smaller hemisphere mass is 
less than $65$~GeV or the larger hemisphere mass is above $105$~GeV.

  Since beam polarisation has been disregarded, W and Z pairs can be 
further diminished due to the dominating $t$-channel production by 
cutting on the polar angle of the thrust axis, $\theta_{\mathrm{thrust}}$. 
Figure~\ref{fig-bmin+cos_thrust} shows that such a cut, although it very 
effectively eliminates the boson pairs, reduces the selection efficiency 
for $q\bar{q}$ events.

  Applying all cuts mentioned, the selection efficiency for 
hadronic $q\bar{q}$ final state is typically $65\%$ while the 
purity is about $90\%$. The efficiency is less than the $85\%$
usually achieved at LEP II, but can be improved by employing the 
beam polarisation option of TESLA or the advantages of high 
resolution vertex detectors.

\section{Observables and corrections}
\label{sec-observables}
The experiments at LEP II routinely use thrust, heavy jetmass,
total and wide jet broadening, $C$-parameter, and the differential
$2$-jet rate from the Durham algorithm to determine $\alpha_S$.
The definition of these observables can be found in 
e.g.~\cite{bib-OPAL-ZPC72-191}. All these
observables are known in second order (${\cal O}(\alpha_S^2)$) and, 
moreover, the leading and next-to-leading logarithms have been resummed 
(NLLA).  Combining these calculations provides a better 
description of the data. This can be expected because the resummed
logarithms dominate in the extreme $2$-jet region while the fixed
order calculation is applicable to the $\ge 3$-jet region. The LEP 
II experiments employ different prescriptions to combine the two 
calculations. For the remainder of this study the $\ln(R)$-matching
scheme is adopted (see e.g.~\cite{bib-OPAL-ZPC72-191}).

  The value of the strong coupling constant is determined from the
differential distributions of these observables. Since this study
relies on Monte-Carlo event generators only, the effect of the
finite resolution and the limited acceptance of a real detector
has to be considered. In \cite{bib-Azuelos-DESY123A} such impacts
were found to be small. 

  To assess hadronisation uncertainties, the approach of \cite{bib-OPAL-ZPC75-193}
is used for this investigation. Some parameters of the PYTHIA program are 
varied to find out the sensitivity of the determined value of $\alpha_S$ 
on the hadronisation correction. The range of variation is given by the 
uncertainties of the parameter values found during the tune of the PYTHIA 
generator. In detail the $b$ parameter of the Lund symmetric fragmentation
is varied by $\pm 0.04/$GeV$^2$ about the tuned value of $0.52/$GeV$^2$. 
The cut-off of the parton shower $Q_0$ is changed from $1.9$~GeV
to $1.4$ and $2.4$~GeV. The width of the gaussian distribution of the 
additional transverse momentum, $\sigma_q$, acquired by the particles 
during the fragmentation is varied by $\pm 0.03$~GeV about $0.40$~GeV. 
Effects of the PYTHIA model of the parton shower development and of the 
fragmentation are investigated using HERWIG and ARIADNE 
alternatively to PYTHIA.

\begin{figure}
\vspace*{-5mm}
\centerline{\epsfxsize=100mm \epsfbox{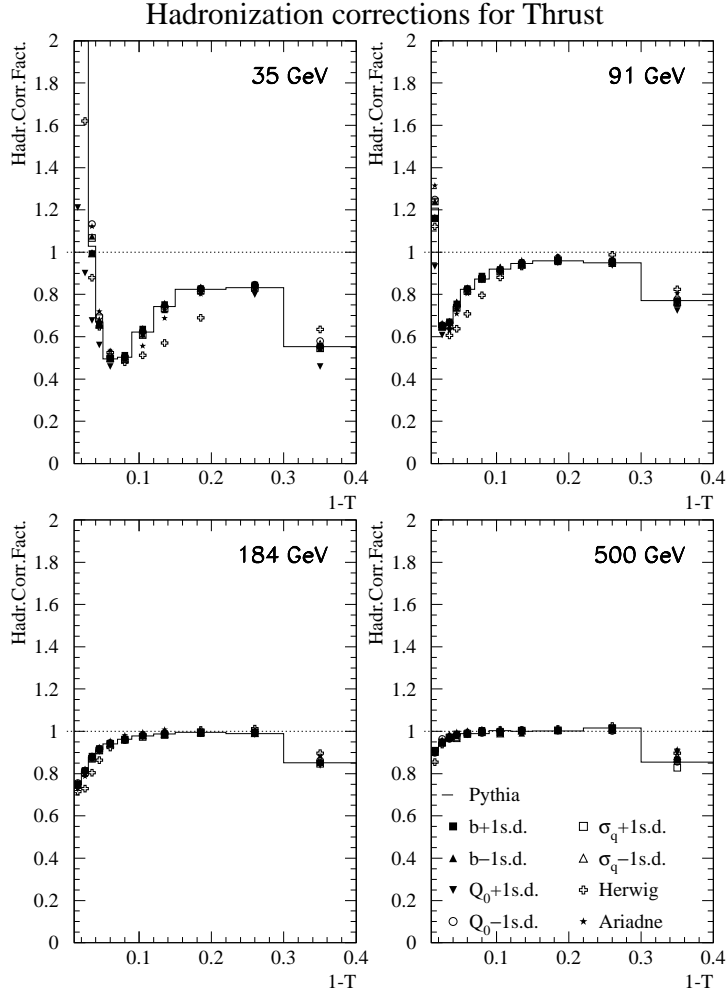}}
\caption{\label{fig-hadronization}
         Hadronisation correction factors for the thrust observable
         at different centre-of-mass energies. The marker 
         symbols indicate different choices of hadronisation parameters
         in the PYTHIA program and other shower and fragmentation 
         models.
        }
\end{figure}
  In general, the hadronisation corrections at a centre-of-mass energy 
of $500$~GeV are much less than $5\%$ in the $\ge 3$-jet region. 
This can be seen from the comparison of the hadronisation corrections 
for the thrust observable in the range of $35$ through $500$~GeV in 
Figure~\ref{fig-hadronization}.
The reduced uncertainty due to the choice of the hadronisation 
parameters at increasing centre-of-mass energies is also noticeable 
from the Figure. 

  Apart from the shrinkage of uncertainties due to the higher 
centre-of-mass energy it should also be kept in mind that a tune 
of the event generators at the end of LEP using the full data
statistics of all experiments will definitely improve the 
precision of the parameters. Thus the range of variation 
will be more constrained than it is for this investigation.
In conclusion, hadronisation corrections and associated
uncertainties will be much reduced at TESLA energies.

\section{Estimation of the error budget of $\mathbf{\alpha_S}$}
\label{sec-alphas}
\begin{figure}
\vspace*{-5mm}
\centerline{\epsfxsize=100mm \epsfbox{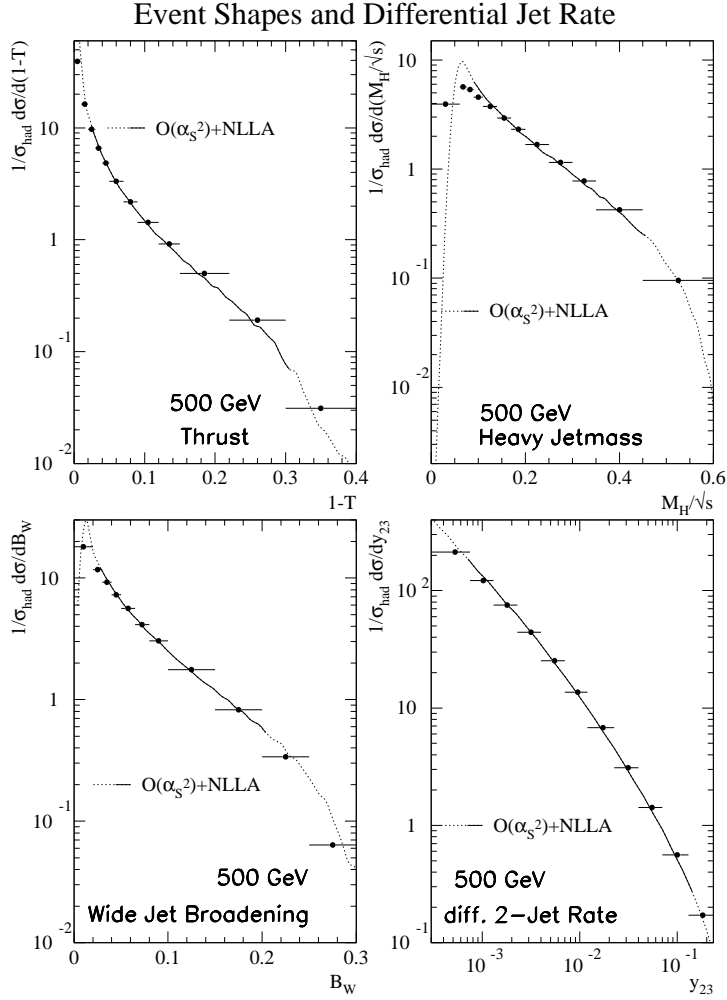}}
\caption{\label{fig-evshapes}
         Event shape and differential $2$-jet distributions 
         fitted using the $\ln(R)$-matched second order and 
         resummed calculations (${\cal O}(\alpha_S^2)+NLLA$). 
         The extrapolation of the fit, whose range is indicated 
         by the solid line, is shown by the dotted line.
        }
\end{figure}
\begin{table}
\begin{center}
\begin{tabular}{|l||c|l|}
\hline
   source & $\delta(\alpha_S(M_{\mathrm{Z}}))$ & comments 
\\ \hline\hline
      statistics    & $\pm 0.0001$ &
\\ \hline
      experimental  & $\pm 0.002 \dots 0.004$    & taken from LEP II
\\ \hline
      hadronisation & $\pm 0.0006 \ldots 0.0015$ & $Q_0$ dominates
\\
      HERWIG        & $+0.017 \ldots +0.025$    & generator not tuned
\\
      ARIADNE        & $-0.004 \ldots +0.002$    &  
\\ \hline
      renormalisation scale
                     & $\pm 0.001 \ldots 0.004$  & 
\\ \hline\hline
      total uncertainty
                     & $\pm 0.004$               & 
\\ \hline 
\end{tabular}
\end{center}
\caption{\label{tab-alphas-error-budget}
        Compilation of the error budget of a determination of
        the strong coupling constant from six observables
        at a centre-of-mass energy of $500$~GeV. 
        The quoted ranges indicate the spread of the $\alpha_S$ 
        values obtained from the different observables.
        }
\end{table}
The error budget of an $\alpha_S$ determination is compiled from 
different sources. Since a perfect detector has been assumed, the 
corresponding experimental uncertainties are adopted from measurements 
at LEP II (e.g.~\cite{bib-OPAL-PN377}). 
Effects due to the roughly   
$10\%$ background from boson pairs are neglected. The statistical 
uncertainty is estimated by considering a $65\%$ selection efficiency 
and an expected integrated luminosity of $300 {\mathrm{~fb}}^{-1}$ 
within one year. This corresponds to about $0.5$ million hadronic 
events which have been generated using the PYTHIA generator employing 
the tuned set of parameters.

  The simulated events are used to assess the remaining uncertainties 
due to hadronisation corrections and due to the choice of the 
hadronisation parameters by fitting the $\ln(R)$-matched 
second order (${\cal O}(\alpha_S^2)$) and resummed calculations 
(NLLA) to distributions of the observables in the hadronic final 
state. Distributions from the thrust, 
heavy jetmass, total and wide jet broadening, $C$-parameter, 
and the $2$ to $3$-jet flip $y_{\mathrm{cut}}$ value, $y_{23}$,
obtained from the Durham jet algorithm are used. Hadronisation 
correction factors are applied to these distributions 
before fitting the theoretical expressions. The result of the
fits is exemplified for thrust, heavy jetmass, wide jet
broadening, and the jet flip in Figure~\ref{fig-evshapes}.
In general good fits are obtained. However, for the heavy jetmass
and the wide jet broadening an excursion of the fitted curve from
the simulated data is visible at small values of these observables. 
This could indicate a problem of either the theoretical formulae
or of the event generator program. The fit must, therefore, not go 
too far into this $2$-jet region, which represents the bulk of
statistics indeed.

  Finally, also the impact of a variation of the renormalisation
scale factor, $x_{\mu}$, is considered. As has become usual $x_{\mu}$
is varied from unity to $0.5$ and $2$, respectively, assigning the
change in $\alpha_S(\sqrt{s})$ as the uncertainty due to the choice 
of the renormalisation scale factor.

 Table~\ref{tab-alphas-error-budget} lists the individual contributions
to the total uncertainty of $\alpha_S$
at the Z mass
scale, $M_{\mathrm{Z}}$.
To calculate a total uncertainty, the individual error contributions
from all six observables are averaged for each systematic
variation individually. The total error is then obtained by adding
the individual contributions in quadrature. Excluding the uncertainty 
derived when choosing the HERWIG generator to correct for hadronisation 
effects, because it is not tuned to the LEP data yet, the total error 
becomes $\delta(\alpha_S(M_{\mathrm{Z}})) = \pm 0.004$. 
This error incorporates a significant contribution of about $\pm 0.003$ due 
to the scale uncertainty. It should be noted that it is of similar size as 
the typical error obtained by the LEP and SLC experiments. The error 
$\delta(\alpha_S(M_{\mathrm{Z}}))$ corresponds to 
$\delta(\alpha_S(500 {\mathrm{~GeV}}))=\pm 0.0025$.

\section{Conclusions}
\label{sec-conclusions}
A feasibility study of the determination of the strong coupling 
constant, $\alpha_S$, at the TESLA linear collider has been 
presented starting from the selection of hadronic events. Neither
initial state radiation nor beamstrahlung constitute a substantial
problem for such a QCD investigation at this collider. 

  The major background stems from the pair production of W and 
Z bosons. Both can be easily suppressed by cutting on the hemisphere 
masses and on the polar angle of the thrust axis. Top quarks 
significantly distort event shape and jet rate distributions 
and must, therefore, also be excluded from an $\alpha_S$ 
determination based on such observables. It has been 
found that a cut on the minimum value of the jet broadening 
effectively removes the contribution from top quarks without 
affecting the selection efficiency. 

  The hadronisation corrections and the associated uncertainties
are very small at TESLA energies. However, the contribution
from different hadronisation models could become a dominating
source of error. 

  Fits of the $\ln(R)$-matched second order and resummed calculations
to six observables have been performed to assess the systematic 
uncertainties of the derived value of $\alpha_S$ using data from 
Monte Carlo generators only. Exerimental uncertainties have been
adopted from the LEP II experiments and are found to be small. 
A very significant contribution to the total uncertainty of
$\delta(\alpha_S(500 {\mathrm{~GeV}}))=0.0025$ stems from the choice
of the renormalisation scale. Its absolute contribution 
will shrink with increasing centre-of-mass energy but the relative
uncertainty is not diminished. To improve on the scale
uncertainty theoretical calculations of third order in $\alpha_S$
are absolutely required.

\appendix
\par
\section*{Acknowledgements}
\par
I would like to thank the organisers of the third ECFA workshop
which took place from November 7-10 in Frascati, Italy.


\end{document}